# An Interactive Automation for Human Biliary Tree Diagnosis Using Computer Vision

Mohammad AL-Oudat [1]   Saleh Alomari [2]   Hazem Qattous [3]   Mohammad Azzeh [4]   Tariq AL-Munaizel [5]

[1] Department of Computer Science, Applied Science Private University, Amman, Jordan,
m.aloudat@asu.edu.jo

[2] Department of Science and Information Technology, Jadara University, Irbid, Jordan,
omari08@jadara.edu.jo

[3] Department of Software Engineering, Princess Sumaya University for Technology,
h.qattous@psut.edu.jo

[4] Department of Computer Science, Applied Science Private University, Amman, Jordan,
m.y.azzeh@asu.edu.jo

[5] Department of Hepatobiliary and Liver transplant, King Hussein Medical Center, Amman, Jordan,
Drtariq2003@gmail.com

**Abstract:**  The biliary tree is a network of tubes that connects the liver to the gallbladder, an organ right beneath it. The bile duct is the major tube in the biliary tree. The dilatation of a bile duct is a key indicator for more major problems in the human body, such as stones and tumors, which are frequently caused by the pancreas or the papilla of vater. The detection of bile duct dilatation can be challenging for beginner or untrained medical personnel in many circumstances. Even professionals are unable to detect bile duct dilatation with the naked eye. This research presents a unique vision-based model for biliary tree initial diagnosis. To segment the biliary tree from the Magnetic Resonance Image, the framework used different image processing approaches (MRI). After the image's region of interest was segmented, numerous calculations were performed on it to extract 10 features, including major and minor axes, bile duct area, biliary tree area, compactness, and some textural features (contrast, mean, variance and correlation). This study used a database of images from King Hussein Medical Center in Amman, Jordan, which included 200 MRI images, 100 normal cases, and 100 patients with dilated bile ducts. After the characteristics are extracted, various classifiers are used to determine the patients' condition in terms of their health (normal or dilated). The findings demonstrate that the extracted features perform well with all classifiers in terms of accuracy and area under the curve. This study is unique in that it uses an automated approach to segment the biliary tree from MRI images, as well as scientifically correlating retrieved features with biliary tree status that has never been done before in the literature.

**Keywords:**  *Biliary Tree; Image Processing; Biliary duct; Deep Learning; Bile Duct; Bioengineering; Bioinformatics.*

---

## 1   Introduction

The biliary tree is a network of vessels that transports digestive juices from the liver, pancreas, and gallbladder to the intestines [1-4]. The most important part of biliary tree is Bile duct. Efferent hepatic innervations is known to play a function in the regulation of bile flow in biliary physiology Innervation may play a role in the modulation of bile salt, cholesterol, and lipid production in response to somatostatin and eating, according to experimental denervation investigations [5].

Obstruction of the biliary tree, which can be caused by stones, tumors that is usually of the papilla of Vater or the pancreas, benign strictures due to chronic pancreatitis or primary sclerosing cholangitis, benign stenosis of the papilla for instance papillary stenosis, or a papillary diverticulum, is the most common cause of dilated bile. Parasites that infiltrate the biliary tree are a frequent type of blockage in impoverished countries. Choledochal cysts, which are either congenital or acquired dilatations without blockage that can be associated with anomalies of the pancreatic duct and the bile duct, are a less common cause of dilated bile ducts [6]. Biliary tree contain several organs intrahepatic, extrahepatic bile ducts and gallbladder. The small bile ducts proximally to the right and left hepatic ducts are the source of intrahepatic cholangiocarcinoma's. The right or left hepatic duct, the cystic duct, or the choledochal duct are all potential sites for extrahepatic bile duct carcinomas [7].

The gallbladder temporarily stores bile produced by the human liver before it is transported via the common hepatic and cystic ducts [8-10]. The gallbladder's main goal is to eliminate bile, which is performed by a hormone called cholecystokinin (CCK), which causes the gallbladder to release bile [11-12]. The CCK causes the gallbladder to contract, allowing all of the bile to flow back out through the cystic duct and on to the common bile duct, which is the final component of the biliary tree [13]. Bile is made up of two components: pile pigments and bile salts, which aid in the emulsification of fat. They allow the body to convert fat into micelles, which are ultimately absorbed in the ileum [14].

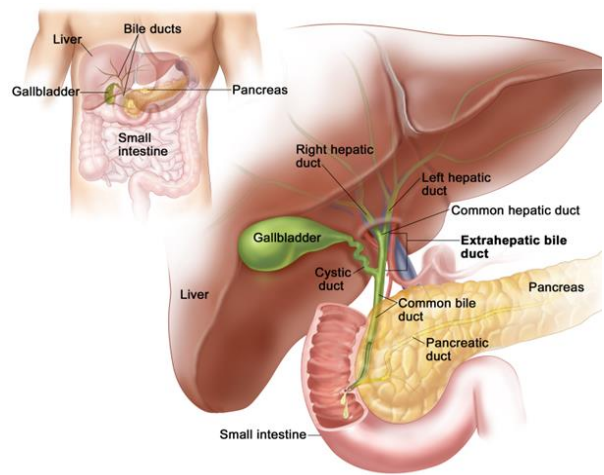

**Fig. 1 Anatomy of biliary tree and bile duct [15].**

As shown in figure 1, that illustrate the anatomy of biliary tree and bile duct. The human biliary tree normal range of healthy bile duct diameter is between 4-8 mm [16]. When the diameter of the bile duct exceeds the usual and healthy range, the bile duct is said to be dilated, which may lead to having a risk of cholangiocarcinoma. The most common cause of dilated bile duct is the obstruction of the biliary tree, which is caused, in turn, by tumors, stones and benign strictures. However, the reasons that lead to the dilated bile ducts are either congenital or acquired dilations without obstruction that can associate with abnormalities of junction between the pancreatic duct and the bile duct. The dilation of the bile duct can be symptoms-free. Therefore, it considers as important sign for doctors to decide that the suspicious patient's biliary tree is dilated or not [17]. However, novice doctors and radiologists find it difficult to recognize the dilation of bile duct. Radiologists are also having difficulty identifying dilation with their naked eyes, particularly if it is less than 3 mm above the usual range. To diagnose the situation, they must obtain an image of the bile duct through Computerized Tomography (CT) scan or Magnatic Resonance Images (MRI) and manually measure the duct's diameter [18]. Although importing the image into the device used for magnifying the image and then physically measuring the diameter is a fundamental and vital operation for diagnostics, it takes a long time. Several research projects have proposed employing visual recognition and feature extraction to aid clinicians and radiologists in recognizing bile duct dilatation. However, just a few of them used MRI pictures in their research. Instead, CT scan pictures were used.

A biliary tree model proposed in this paper in order to help radiologists and doctors to detect and accurate diagnose if the suspected patient may have risk or not. The proposed model using computer vision and machine learning algorithms applied on MRI images to classify the biliary tree status. Moreover, ten features are extracted from the segmented biliary tree which are minor axis, major axis, bile duct area, biliary tree area, compactness, and texture features (contrast, mean, variance and correlation). All these features extracted to help specialists in detecting the status of biliary tree (dilated or not) in order to give an efficient diagnosis.

The proposed model is consisting of three stages. Preprocessing of the image consider as the first stage that resizing the MRI image, then, a gray scale and sharpen technique applied on it [19-20]. After that, the image is denoised using Convolution Neural Network (CNN) [21]. The second stage is segmentation, that uses Contour Without Edges (CWE) technique on the improved MRI image from the preprocessing stage [22]. The purpose of the segmentation stage is to segment the biliary tree. The last stage is extracting ten features from the segmented area. Once all ten features are extracted from all MRI images, multiple of common binary classifiers such as, K-Nearest Neighbors (KNN), Support Vector Machine (SVM), Multi-Layer perceptron neural network (MLP) Logistic Regression (LR), Decision Tree (DT) and Random Forest (RF) are applied to construct the biliary tree dilation detection models.

The MRI images used in this paper were collected from King Hussein Medical Center (Amman-Jordan). The dataset is split into two parts. The first part contains half of the MRI images labeled as normal bile ducts. The other half labeled as dilated bile ducts. The labeling process applied by specialists in the King Hussein Medical Center. Fortunately, the quality of these images is accepted to carry out this research.

To that end, this research attempts to answer the following research questions:

RQ1. Is it possible to detect early biliary tree dilatation using machine learning algorithms?

RQ2. Do the extracted features enhance the diagnosis of biliary tree dilation in comparison to traditional way?

In summary, the main contribution of the proposed research can be specified as follows:

- Proposing a new framework that help both doctors and radiologist in the initial diagnosis of biliary tree dilation, as it the most important indicator for one or more several diseases that the patients may have.

- Introducing a vision-based architecture and features for initial screening using MRI images. This work shows that there is an evidence of computational relationship between the status of biliary tree dilation and the extracted features.

## 2 Literature Review

Nowadays, the dilation of biliary tree becomes one of the most common diseases that can be caused due to many reasons. Bile dilation is difficult for doctors and radiologists to detect. To execute a manual technique for determining the duct diameter, specialists will need CT or MRI scans. Many academics have proposed numerous studies for biliary dilation recognition to tackle these challenges. However, there is a lack of studies regarding bile duct dilation using MRI images as most of the studies are focused on CT images. Medical photos are believed as a valuable resource that can provide invaluable information for medical decisions and scenarios, particularly when used in conjunction with deep learning techniques [23-24]. As a preprocessing step for MRI of the biliary tree, it is critical to reduce any type of noise, as this would obstruct illness diagnosis [25]. [26] Since it analyzes the most recent evolving techniques in the field of medical image processing, notably those techniques connected to the de-noising stages with deep learning approaches, it has been suggested that a new era of picture-based disease diagnosis is beginning. Several deep learning approaches have been used to quantify bile duct dilation in a variety of professional studies. Neural Networks are pioneering approaches in the field of machine learning, and they have a variety of applications in image processing, ranging from backpropagation neural networks to Convolutional Neural Networks (CNN) [27-28]. Exciting proofs and interpretations about the behavior of neural network weights in general, and backpropagation networks in particular, were presented by certain study [29-30]. An important study regarding the uses of neural networks in feature extraction and feature selection was presented in another paper [31]. In contrast of, the CNN machines' performance in image applications, particularly in image de-noising, was impressive. [32] offered an overview of deep learning, stressing the critical importance of deep learning approaches in radiology-based medical practices. The De-noising CNN (DnCNN) is one of these approaches. We discovered that none of the existing MRI image techniques, specifically for bile duct measurement, used the DnCNN to remove any noise in the MR images. The DnCNN, on the other hand, has been extensively used with MRI images of different human organs [33-35]. As a result, the method used in this study is the first to use DnCNN to eliminate noise from bile duct MR images. [36] presented a useful study on how to evaluate bile duct diameter using Magnetic Resonance Cholangiopancreatography (MRCP) Images. The authors of [37] used a method to test the DnCNN's efficiency by introducing some artificial noise to be rectified, and the results showed extremely modest changes. Furthermore, [38] developed a new deep learning de-noising technique based mostly on two steps. The liver and bile tree were treated using a variety of treatments and techniques. Ultrasound pictures, Computed Tomography (CT), MR Cholangiopancreatography (MRCP), Magnetic Resonance Imaging (MRI) and other types of images were used in these approaches, etc. Though, the magnetic resonance imaging (MRI) is a developing technique that is generally acknowledged and used in the medical treatment of biliary illnesses [39]. The Contour Segmentation Technique (CST), which measures the contour, volume, and shape of the liver or bile tree, is one of the most prevalent procedures used in this approach. In this study, we employed the Contour Segmentation Technique to measure the intrahepatic biliary channel diameter, a function for which the Contour Segmentation Technique has never been used before. In [40], the CST technique was used in conjunction with 4D models of liver anatomy to ensure adequate surgical resection and avoid any unique anatomy related to the biliary information needed during surgery. CT and MRI images do not provide this type of information. As a result, 4D models are thought to be the most suited technique. Nonetheless, the strategy provided by [41] is also regarded as an effective technique for computer assisted surgical planning that uses 3D imaging. [42], on the other hand, proposed a unique method for precisely measuring liver volume using MRI. The contour segmentation method was the primary technique used to define the liver's boundaries. Similarly, [43] discussed the impact of MRI-based techniques and their evolving importance in data acquisition and processing. They discovered that MRI had made significant progress in the fields of research and medical treatment procedures relating to the biliary duct in particular. Similarly, to what [43] proposed, [44] proposed an approach to examine the biliary duct using MRI and MRCP images. Other studies used ultrasound-based techniques on the extrahepatic bile duct to detect possible duct dilatation and, as a result, diagnose primary neuroendocrine [45] and biliary duct assessment [46]. The proposed framework in this study, and according to the related work review, will be the first to use the DnCNN technique in combination with the CST to handle the intrahepatic biliary ducts from MRI images and extract a useful set of features and values that expresses the diagnostic condition of this organ.

## 3 Methods and Experiments

The major goal of this research is to create an MRI-based automated biliary tree dilation detection framework. As illustrated in Figure 2, the proposed framework includes a series of processes for processing, segmenting, and extracting features from MRI images in order to create training data that best describes the biliary tree. This information will be fed into a machine learning algorithm that will be used to train and test a prediction model that can forecast a patient's condition based on their MRI scans.

### 3.1. Dataset

In this research. The MRI images dataset[1] used in this proposed work was collected from The King Hussein Medical Center's (Amman-Jordan). For each patient, 100 images were randomly selected from the database. The total number of images collected and adopted for this research was 100 images. the dataset split into two parts by medical experts as "having dilated biliary tree" or "having normal biliary tree". Samples of 50 participants were collected randomly from each group. The collected samples contained a total of 43 female and 57 male participants. The age range of the participants' is 40 – 65 years old. Further examination for each image was conducted by an expert (surgeon) to make sure that there is no mistake during the initial classification. The surgeon stated the dilation level of each bile duct in the images; and accordingly, the bile duct in each image was labeled as Normal or Dilated taking into consideration that the diameter of a normal bile duct ranges from 4mm up to 8mm [16].

---

[1] IRB approval has been obtained at King Hussain Medical Center for the human subject samples. The authors formally requested anonymous access to the dataset.

### 3.2. Preprocessing

This is the initial step of the proposed framework, as shown in Figure 2. It begins by passing the MRI image through a cascade preprocess technique. The MRI image is resized to 512×512 scale in the first step of the pre-processing phase. This scale was chosen to preserve the original resolution at a reasonable level of quality. The image is then converted grayscale to be sharpened. as shown in the equation 1:

$$Grayscale\ Image\ = 0.3 \times Red + 0.59 \times Green + 0.11 \times Blue \tag{1}$$

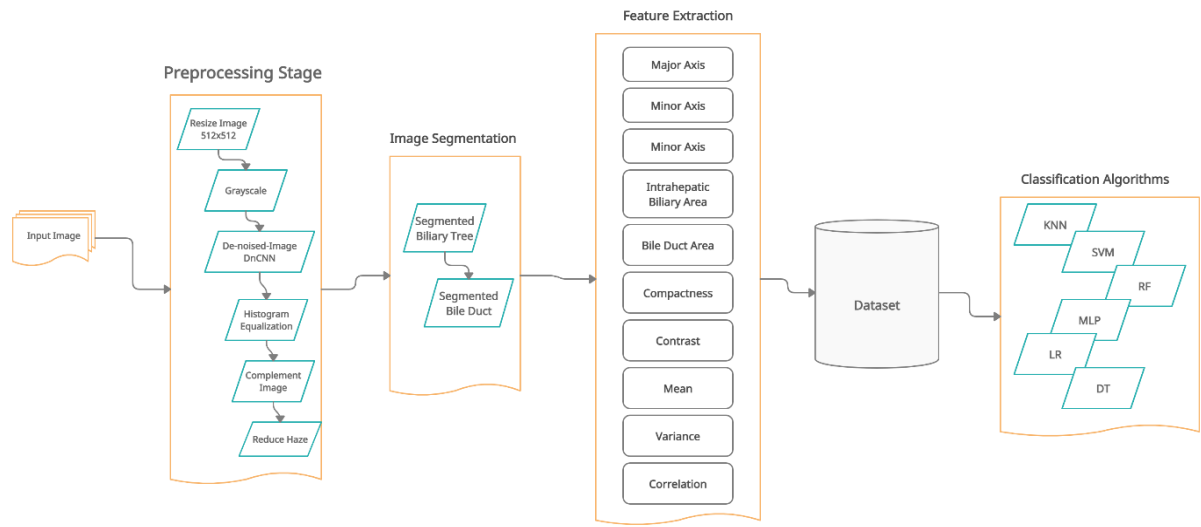

**Fig. 2 The proposed framework.**

To remove any existing noise, the produced gray scaled image is sent into the trained neural network. Feed-forward De-noising Convolution Neural Networks are used for this (DnCNN) [21] was used, as shown in Figure 3.b.

This method is mostly used to deal with Gaussian de-noising at any noise level. The latent noise-free picture in the DnCNN's heading layers is implicitly removed by DnCNN. In a nutshell, DnCNN divides the image into many stages:

- Set the convolution filter size to $3 \times 3$.
- Set the receptive field size to $35 \times 35$ and applies Gaussian de-noising.

Apply residual learning to train residual mapping based on three layers (Convolution layer, Relu Layer and Batch Normalization layer). Finally, DnCNN trained in an end-to-end fashion based on Equation 2.

$$l(\Theta) = \frac{1}{2N}\sum_{i=1}^{N} \| R(y_i; \Theta) - (y_i - x_i) \|_f^2 \tag{2}$$

Where ($\Theta$) is the trained parameters, N represents the number of images, $x_i, y_i$ represent the noise-free image for training and $R(y_i; \Theta)$ represents the residual learning [21]. Next, the process of image enhancement is initiated via histogram equalization which is considered as a technique for contrast adjustment that provides better image quality without any information loss [48], as Figure 3.c shows. Equation 3 below. illustrates the histogram equalization technique.

$$h(v) = round(\frac{CDF(v) - CDF_{min}}{(M \times N) - CDF_{min}} \times (L-1)), \tag{3}$$

where CDF is the cumulative distribution function, $CDF_{min}$ is the minimum non-zero value of the cumulative distribution function, M and N are the width and the height of the image respectively, and L is the number of the used gray level.

After that, an image complement technique is used to prepare the MRI image for the next enhancement, which uses the De-Hazing approach to distinguish the biliary tree from the surrounding component of the picture. The atmospheric haze is reduced using the De-Hazing process. It all starts with a calculation of the ambient light. The transmission map is then estimated and refined, and the image is finally restored [49-50]. The resultant MRI image after performing the De-Hazing approach is shown in Figure 3.c. Figure 3.d depicts the final stage of the preprocessing phase, which entails complementing the image once again in order to prepare it for the second phase, which is the segmentation phase.

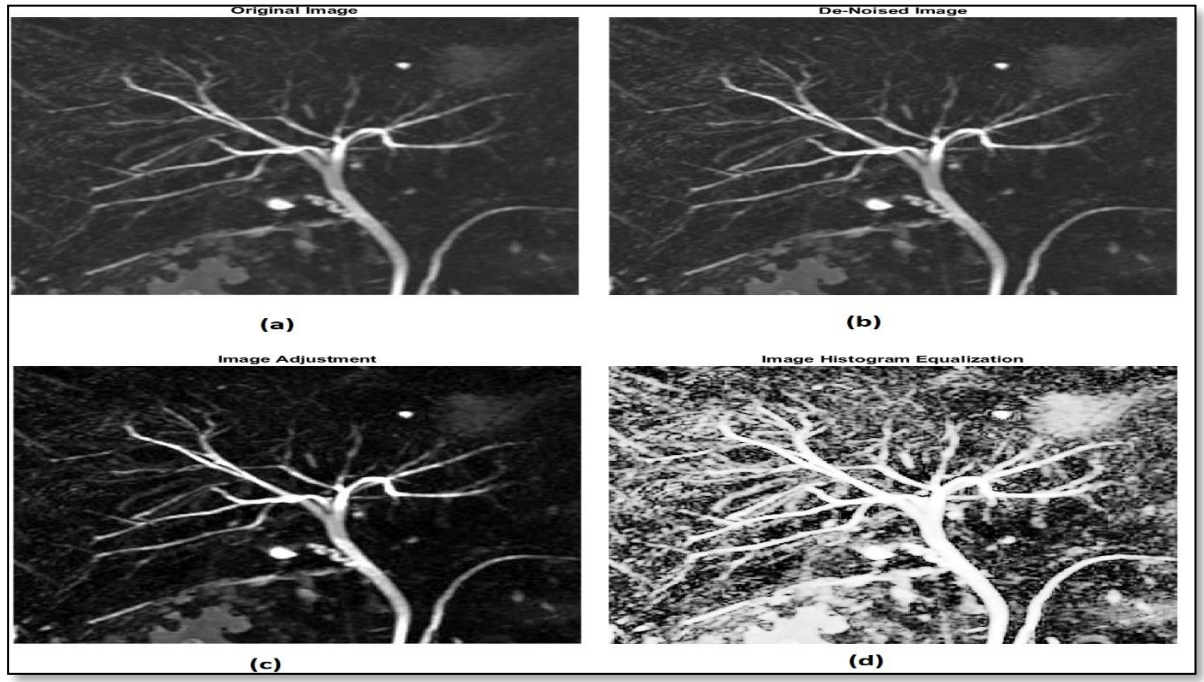

**Fig. 3 Pre-Processing techniques. (a) Original Image. (b) De-noised image using DnCNN. (c) Image adjustment and then histogram equalization. (d) Complemented and reduced haze MRI Image.**

### 3.3 Segmentation

The image is ready for segmentation once the preparation phase is completed correctly. For the segmentation process, the enhanced active contour technique was used [22]. This technique has the advantage of defecting objects with undefined bounds using a gradient or smooth boundary. Furthermore, the enhanced active contour approach can work even if there is still considerable noise in the image [22]. The segmentation procedure begins with the generation of a black mask image of the preprocessed image's size. Figure 4.a shows the white rectangle inside the mask image that serves as the process's beginning point, also known as a seed for the segmentation approach used. The biggest issue we have while making the mask that carries the seed is determining the best location to put the seed. As a result, we conducted multiple trials and discovered that the exact spot to plant the seed is 10 pixels to the left of the MRI picture center. Equations 4 and 5 are used to determine the placement of the seed.

$$x = \frac{W}{2} - 10, \qquad (4)$$

$$y = \frac{H}{2} - 10, \qquad (5)$$

where W, H are the width and the height of the mask image respectively.

As the initial segmentation point for all images, this location is assured. Identifying the exact position of the seed, on the other hand, only solves part of the problem. The other element is solved by choosing the proper seed size. The seed size is significant because, in certain situations, the seed does not reach the bile duct adequately, and so the segmentation process does not begin. We changed the size of the seed to solve this problem. Equations 6 and 7 show the coordinates of seed adapted size.

$$x_i = \frac{r_i}{2} - 10, x_2 = \frac{r_i}{2} + 10, \qquad (6)$$

$$y_i = \frac{c_i}{2} - 10, y_2 = \frac{c_i}{2} + 10, \qquad (7)$$

Where $r_i$ is index of ith row, and $c_i$ is the index of ith column.

Figure 4.a depicts the seed placed in the precise location and size (the white rectangle). The segmentation is now ready to be implemented. The adopted contour technique segmentation works by involving several iterations over the image starting from the predefined seed point. When a predetermined number of iterations have been accomplished, the segmentation process comes to an end. The key need of the enhanced active contour technique is to know how many iterations are required. As a result, various studies were carried out using MRI pictures of the dilated and normal biliary tree. We discovered that 625 iterations are required. Equation 8 describes how the active contour approach works. The segmentation process is shown in Figure 4.b at iteration number (60, 340, 500 and 625). The detected biliary tree in the image under examination is the principal result of this step, as shown in Figure 4.c.

$$F(c_1, c_2, C) = \mu \times |C| + v. Area(inside(c)) + \lambda \int |u_0 - c_1|^2 dxdy + \lambda \int |u_0 - c_2|^2 dxdy, (8)$$

Where C represents the curve of the biliary tree, the constants $c_1, c_2$ are the averages of $u_0$ inside and outside C respectively, $u_0$ represents the image gradient, and $\mu, \lambda$ are two positive parameters.

### 3.4. Features Extraction and Selection

Once the segmentation process of biliary tree is accomplished, then, the process of segment the bile duct can started. Binary Large Object (BLOB) technique is adopted in order to accomplish this. The perimeter and circularity of the blob technique are the two most important characteristics. As a result of using the blob technique, multiple blobs are formed from the segmented biliary tree, and the bile duct is recognized as the largest linked component between the clustered blobs. The extracted bile duct from the biliary tree image is shown in Figure 4.d. The diameter of the bile duct can be determined by measuring the distance between two places on the width and height of the bile duct. As a result, these two features were chosen for recognition and given the names Minor Axis (MIA) and Major Axis (MJA). Figure 4-d. shows the MJA and MIA, which represent the horizontal and vertical distances between pixels in the bile duct, respectively. Furthermore, because the normal bile duct has a smaller area than the dilated bile duct, the Bile Duct Area (BDA) is regarded as a significant element in medical diagnosis. We also discovered that the biliary tree structure of the dilated bile duct provides more information than the structure of the normal bile duct. The right and left hepatic ducts, in instance, look sharper in MRI images of a dilated bile duct than in a healthy bile duct [51]. As a result, the Intrahepatic Biliary Area (IBA), also known as the biliary tree area, is considered as a feature. Compactness (CMP) is also a critical property because it determines how tightly pixels are packed together. The compactness is seen in Equation (9).

$$Compactness = \frac{P^2}{4\times\pi\times a}, \qquad (9)$$

Where $P$ represents the perimeter, and a represents the area. The image in figure 4.e shows the difference in structure between the normal and the dilated bile ducts.

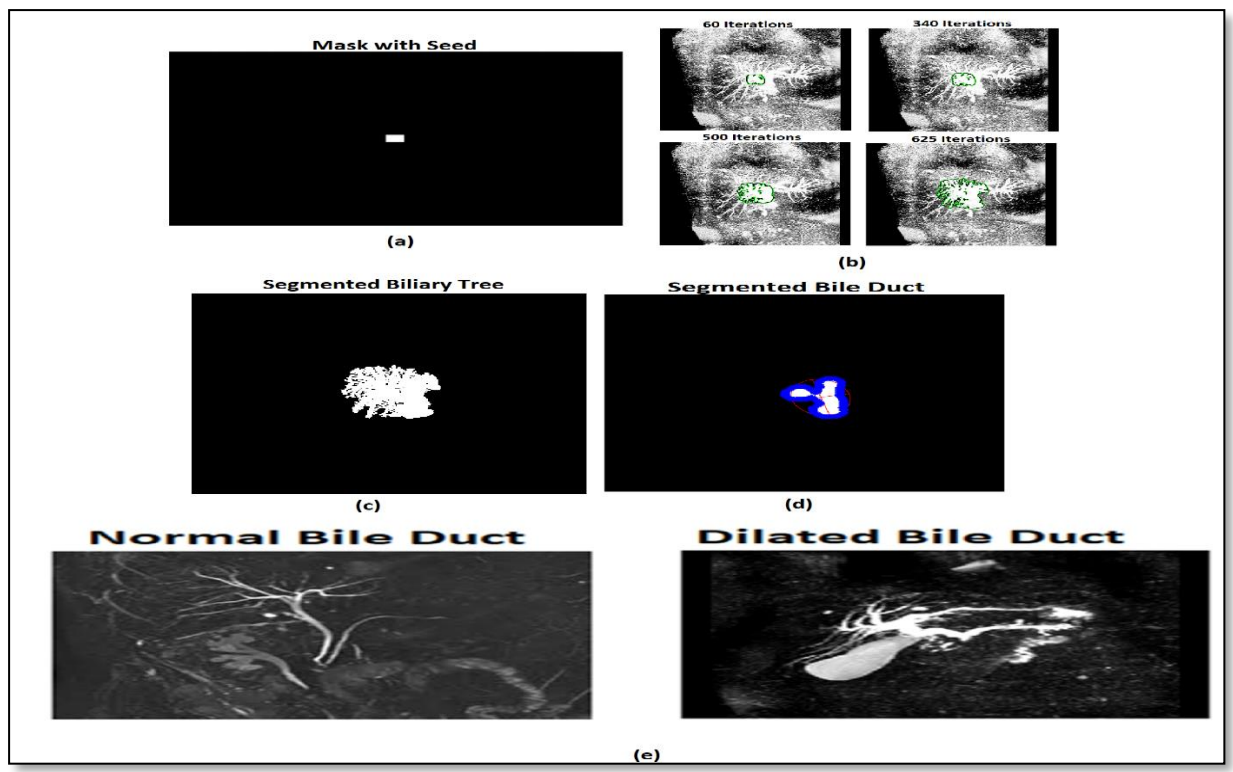

Fig. 4 Experimental and feature selection. a Mask Image. b Active Contour Segmentation on 60, 340, 500 and 850 Iterations. c Segmented biliary tree. d Major Axis and Minor Axis of Segmented Bile Duct. e Structure of Normal and Dilated Bile Ducts.

The texture feature categories are GLCM is a tabulation of how often different combinations of gray levels co-occur in an image or image section, can provide useful information about the texture of an image but cannot provide information about shape. GLCM specifies the number of times that the pixel with value (i) occurred horizontally adjacent to a pixel with value j [52]. After creating the GLCMs using gray comatrix, it can derive several statistics from them include Contrast (Cont), Correlation (Corr) [53]. Contrast that a measure the local variations in the gray-level co-occurrence matrix. The mathematical formula to calculate Contrast as shown in Equation 10. ($p_{i,j}$) represents the element of GLCM matrix.

$$\text{Contrast} = \sum_{i,j=0}^{N-1} p_{i,j}(i-j)^2 \tag{10}$$

Correlation that a measure the joint probability occurrence of the specified pixel pairs and mathematically started by calculate each mean and the variance (Var) by the Equations 11 and 12 then calculate the Correlation by the Equation 13.

$$\text{Mean}(\mu) = \sum_{i,j=0}^{N-1} i\, p_{ij} \tag{11}$$

$$\text{Variance}(\sigma^2) = \sum_{i,j=0}^{N-1} p_{ij(i-\mu)^2} \tag{12}$$

$$\text{Correlation} = \sum_{i,j=0}^{N-1} \frac{(i-\mu)(j-\mu)}{\sigma^2} \tag{13}$$

However, all of the extracted features are measured in pixels, which will have an impact on the classification results due to the various sizes of the original images. As a result, we take the ratio into account for features that use pixels as a unit. As a result, the major and minor axes are each divided by 512 (Which is the size of the image). Furthermore, we divide the bile duct area by the biliary tree area to provide a feature termed Area Ratio (AR). As a result, the feature set is pared down to only four: MJA, AR, MIA and CMP. A simple comparison of the average selected feature values between the dilated biliary tree and the healthy biliary tree was done to determine the initial efficiency of the selected features. The final data description for all retrieved features is shown in Table 1. To ensure that all features have the same influence when applying machine learning techniques, each feature was scaled to be within the range [0, 1] using the min-max scaler algorithm.

Table 1  The complete dataset description after extracting the features and before Scaling.

| Medical | MJA | MIA | BDA | BTA | CMP | AR | Cont | Var | Corr | Mean |
|---|---|---|---|---|---|---|---|---|---|---|
| mean | 9.78 | 132.69 | 61.49 | 5854.69 | 9171.73 | 12.48 | 0.10909 | 0.07791 | 0.947737 | 0.038823 |
| std | 4.43 | 58.58 | 36.05 | 6955.09 | 8365.57 | 8.13 | 0.2021 | 0.02051 | 0.039886 | 0.031791 |
| min | 4.00 | 30.54 | 26.49 | 697.00 | 820.00 | 0.72 | 0.0038 | 0.02896 | 0.68176 | 0.002785 |
| Max | 18.00 | 303.35 | 214.53 | 44843.00 | 41982.00 | 33.09 | 0.96480 | 0.11074 | 0.99285 | 0.140656 |

### 3.5 Choice of Machine Learning Algorithms

In this study, Multiple machine learning algorithms, such as Multi-Layer Perceptron Neural Network (MLP), Support Vector Machine (SVM), K-Nearest Neighbors (KNN), Logistic Regression (LR), and Random Forest (RF), are used to build a classification model that can predict whether patients are dilated or normal biliary tree.

A feed-forward neural network technique with one input layer, at least one hidden layer, and one output layer is known as the MLP. An input vector is represented by each neuron in the input layer. This model's input variables are the four features (MIA, MJA, AR, and CMP) from the preceding section's reduced feature list. In the neurons of the buried layer, the MLP employs a nonlinear activation function. In the output layer, on the other hand, a linear activation function is frequently utilized. The number of neurons in the output layer is determined by the type of problem. The number of neurons equals the number of labels, and the output is the probability of each label if the problem type is classification, as it is in our instance. Patients' status as one possible output label from dilated, normal is the target variable. As a result, each output neuron represents a class label, and the output neuron with the highest probability is chosen. The number of hidden layer neurons vary depending on the number of input neurons and the training algorithm utilized.

The backpropagation method and the conjugate gradient algorithm are two popular training algorithms. We used the backpropagation technique in this work because it has advantages over the conjugate gradient algorithm. After numerous experiments, the number of neurons for each layer was carefully selected. The number of input neurons is 4, which corresponds to the number of input characteristics, the number of hidden neurons in the hidden layer is 10, and the number of output neurons is 2. The sigmoid function was chosen as the activation function in this study.

To classify data, the KNN employs the concepts of retrieving by similarity and voting. The KNN basically finds the closest k comparable examples for the new one, then uses voting to determine the final output. Choosing the value of k has a significant impact on KNN accuracy; for example, if we choose a little value of k, more valuable examples may be overlooked, reducing accuracy, whereas a big value of k is time and resource intensive. The most frequent method for determining the proper value of k is to multiply the total number of data points by the square root of the total number of data points. We chose k=5 in this article because it is a sensible figure that allows us to select the best closest scenarios while minimizing resource costs.

SVM is used to create the best hyperplane for separating data with the most margin. The margin is defined as the maximum width of the slab parallel to the hyperplane that has no data points on the inside. The choice of kernel functions, such as Gaussian, Polynomial, and Radial Basis Function, determines the best hyperplane generation. Hyperplane generation can benefit from both Gaussian and Radial Basis function kernels because they support the localization of training data, which implies that the data can be efficiently separated. Radial basis kernel was employed in this research.

Random Forest is an ensemble learning technique that may be built using the Bagging process from a set of decision trees. While expanding the trees, RF adds more unpredictability to the model. When splitting a node, it looks for the best feature from a random subset of features rather than the most essential feature. As a result, there is a lot of variety, which leads to a better model. As a result, the process for splitting a node in a random forest only considers a random subset of the features. Instead of searching for the greatest possible thresholds, you may make trees even more random by employing random thresholds for each feature (like a normal decision tree does).

LR is a binary classifier based on the probabilistic statistical regression technique's assumptions. By fitting cases to a logistic curve, LR is used to describe the connection between one dependent binary variable and one or more independent variables. The sigmoid function is used by the LR to estimate the class probability of a test item.

### 3.6 Performance Measures

The proposed framework's accuracy has been assessed using a variety of performance measures. Sensitivity, precision, specificity, F1 score, accuracy, and Area Under Curve were the major performance measurements we used (AUC). As stated in equation 14, sensitivity is utilized to calculate the proportion of dilated cases that are anticipated to be dilated. As illustrated in equation 15, precision is utilized to calculate the fraction of dilated patients from total dilated forecasts. As indicated in equation 16, specificity is utilized to determine the fraction of normal instances that are anticipated to be dilated patients. As indicated in equation 17, F1 is utilized to make choices between sensitivity and precision.

As indicated in equation 18, the accuracy measure is used to assess the classification model's overall accuracy. Finally, AUC calculates the area under the ROC curve, which is generated by a set of Sensitivity and Specificity values and expressed as a single value between 0 and 1.

$$Sensitivity = \frac{TP}{TP + FN} \qquad (14)$$

$$Precision = \frac{TP}{TP + FP} \qquad (15)$$

$$Specificity = \frac{TN}{TN + FP} \qquad (16)$$

$$F1\ score = 2 \times \frac{(sensitivity \times precision)}{(sensitivity + precision)} \qquad (17)$$

$$Accuracy = \frac{TP + TN}{TP + TN + FP + FN} \qquad (18)$$

Where TP: True positive, TN: True Negative, FN: False Negative and FP: False Positive

## 4 Results and Discussion

In this research, we developed an automated framework to aid doctors and radiologists in the initial screening of the biliary tree's state. We used numerous image processing algorithms in the proposed framework to segment the biliary tree and the bile duct from MRI data. Ten features are collected from each MRI picture once the region of interest has been segmented. Major Axis (MJA), Minor Axis (MIA), Intrahepatic Biliary Area (IBA), Bile Duct Area (BDA), Compactness are the characteristics (CMP), and some texture features such as contrast, mean, variance and correlation. By dividing IBA by BDA, a new feature called Area Ratio (AR) has been created. As a result, the IBA and BDA were not included in the prediction models that were created. We feel that the dataset acquired from all images is ready to be sent to the neural network classifier after applying the ratio concept to the extracted features. Following the extraction of the desired feature from all MRI images, an analysis was performed to show the data distribution between the retrieved features and clinical measurements. The link between the size of the biliary tree volume in millimeters (Medical) and each attribute is shown in Figure 5. Experts utilize the medical variable to identify whether or not there is a possibility of biliary tree dilatation. Dilated individuals have blue points, while normal patients have orange points. The graphic clearly shows that normal patients have a smaller biliary tree volume than dilated patients, indicating that there are significant variations in all extracted features between normal and dilated biliary tree features. These findings suggest that the derived features will aid in the first screening of the biliary tree's condition. This is because these characteristics can distinguish a dilated biliary tree from one that is healthy. Figure 6 depicts the boxplots of each variable after scaling, grouped by class label value, dilated (Yes), and normal (No). First, we used min-max scaling to ensure that all features had the same influence and to make comparisons between them easier. The boxes and median values are not completely overlapped in the figure, indicating that there is a considerable discrepancy between two labels. Surprisingly, the median of all traits with normal patients is lower than the median of dilated patients. This is a measure of how well the extracted features can tell the difference between two class labels. Over the Biliary Tree area variable, we can see that the boxes and median values of two labels do not overlap. Interval graphs in Figure 7 corroborate these findings. We can observe that the interval plots for two class labels in each characteristic are not completely overlapping, indicating that significant differences exist between two groups of patients.

The gathered data is fed into machine learning algorithms. We performed 10-folds cross validation to analyze our outcomes for this study. The entire dataset was partitioned into ten subsets, with one subset serving as a test and the remaining subsets serving as training for the classification models in each validation run. Each run's evaluation measures are recorded, and average values are calculated. Table 2 shows the aggregated evaluation results. We can see that all models perform admirably on all evaluation criteria. In terms of AUC, which many researchers use to assess the accuracy of classification models, we can see that all models have high area under curve values, implying that the models can accurately categorize dilated patients based on our extracted features. The SVM is the better of the two, as indicated in bold language in Table 2. Other performance indicators suggest that RF is the best option. Another key point to note in this study is the classification algorithm's ranking consistency across datasets and numerous performance indicators. The algorithm's ranking stability means that it should always produce accurate results and rank high across all datasets and performance indicators. The RF was ranked best across all evaluation measures except the AUC measure, according to our findings. Finally, we may deduce from the aforementioned findings that the KNN is the least accurate across all performance measures. The ROC for all models is shown in Figure 8. The RF is clearly superior, which validates the findings in Table 2, whereas the LR is the less accurate.

Again, We go over the study questions again to see if we can answer these questions based on our findings:

**RQ1.** Is it possible to detect early biliary tree dilatation using machine learning algorithms?

Yes, we discovered that all classification models generate highly accurate results with good performance on the dataset derived from the retrieved features. Figure 8's ROC backs up our claim that all created models are predictive.

**RQ2.** Do the extracted features enhance the diagnosis of biliary tree dilation in comparison to traditional way?

Yes, as can be shown in Figures 5, 6, and 7, the retrieved characteristics are capable of distinguishing between two class labels (dilated and normal). Furthermore, all classification models' accuracies corroborate our hypothesis.

Table 2  Evaluation results of all machine learning models.

| Model | Accuracy | Sensitivity | Precision | F1 Score | Specificity | AUC |
|---|---|---|---|---|---|---|
| KNN | 0.88 | 0.91 | 0.88 | 0.90 | 0.89 | 0.96 |
| SVM | 0.91 | 0.84 | 0.94 | 0.89 | 0.97 | 0.97 |
| RF | 0.97 | 0.94 | 0.98 | 0.95 | 0.99 | 0.95 |
| MLP | 0.92 | 0.85 | 0.97 | 0.88 | 0.97 | 0.93 |
| LR | 0.86 | 0.81 | 0.95 | 0.87 | 0.95 | 0.91 |
| DT | 0.95 | 0.90 | 0.89 | 0.89 | 0.96 | 0.90 |

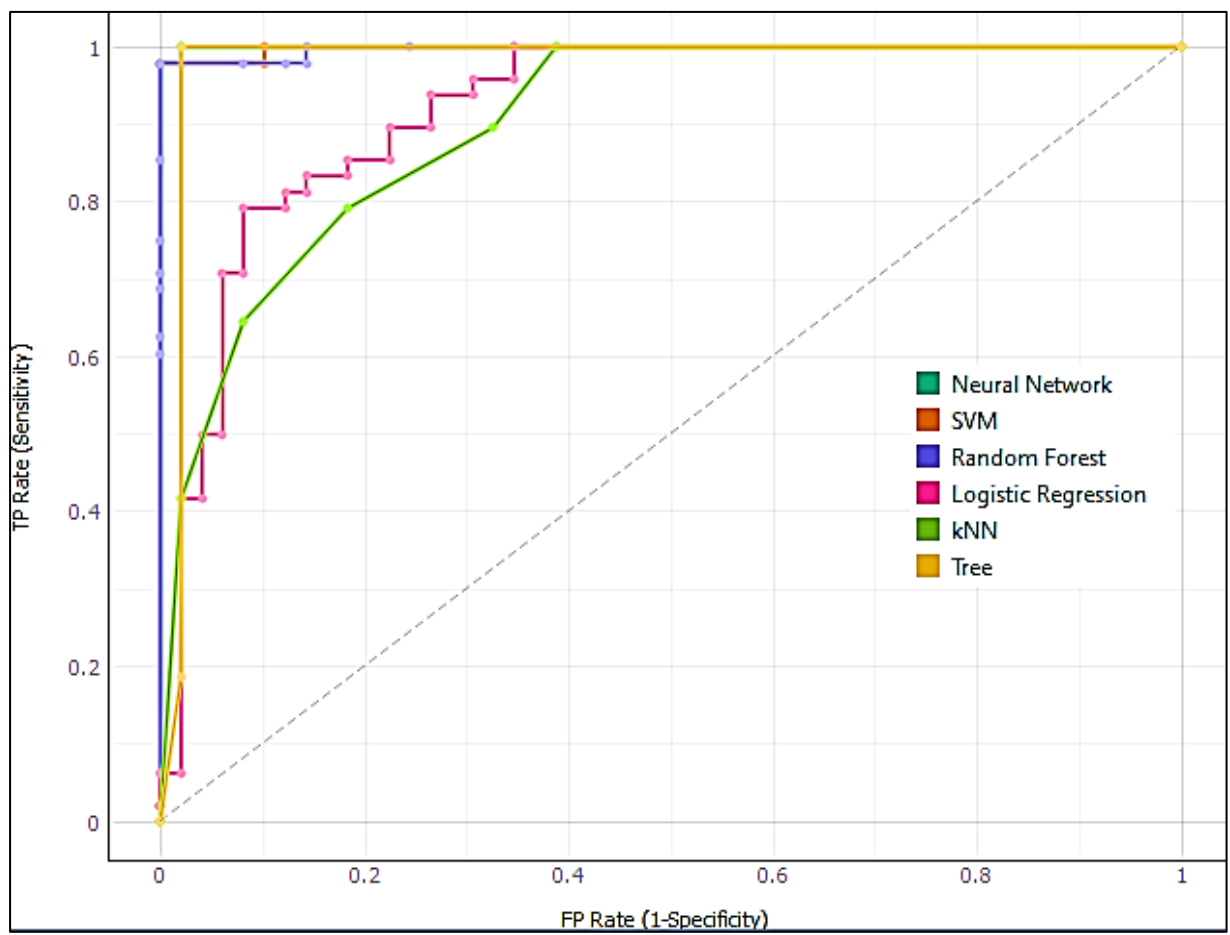

Fig. 7 ROC graph for all machine learning models.

## 5  Conclusion

In this paper, a novel automated framework for biliary tree MRI imaging has been presented to aid clinicians and radiologists in the first screening of the biliary tree. To aid in the detection of biliary tree dilatation, vision-based technologies such as image processing and machine learning techniques have been explored. The MRI pictures were obtained at King Hussein Medical Center (Amman, Jordan), with the first half of the images representing normal bile ducts and the second half representing dilated bile ducts. Specialists in the medical field labeled each MRI image. The biliary tree was segmented using a variety of image processing techniques. The features extraction process begins when the region of interest has been split. Major Axis (MJA), Minor Axis (MIA), Intrahepatic Biliary Area (IBA), Bile Duct Area (BDA), and Compactness were all retrieved from the biliary tree (CMP). After the features extracted, they were subjected to a ratio technique in order to remove the units (pixels). The features were then stored in the dataset and fed to six binary classifiers. The results reveal that the retrieved features let classifiers discriminate between dilated and normal patients perfectly. Furthermore, it has been discovered that all classification models give valid predictions when tested on a dataset derived from extracted features. More MRI images could be added to the database in the future to train the framework on more scenarios. More vision techniques, such as deep neural networks for segmentation, can be used to MRI pictures in the future, and more classifiers can be added and compared to determine which classifier has the best accuracy.

Finally, it's crucial to note that our research is fraught with difficulties. One of the difficulties is the data collection procedure, which is made more difficult by the fact that there is no online dataset available for research purposes. As a result, gathering data from the hospital requires too much effort and time.

**Acknowledgements**

The authors are grateful to the Applied Science Private University, Amman-Jordan, for the full financial support granted to cover the publication fee of this research article.

The authors would like to take opportunity to thank King Hussain medical center for their support and grant permission to collect MRI images.

# References

[1] P. Frédéric, Lemaigre. *Development of the Intrahepatic and Extrahepatic Biliary Tract: A Framework for Understanding Congenital Diseases*, Annual Review of Pathology: Mechanisms of Disease, Volume 15:1, pp. 1-22, 2020.

[2] N.C. Hul, G.R. Schooler, E.Y Lee. *Bile Duct and Gallbladder*. Pediatric Body MRI, Springer, Cham 2020.

[3] L. Centonze, S. Di Sandro, I. Mangoni, L. De Carlis. *Biliary Reconstruction Techniques: From Biliary Tumors to Transplantation*, Liver Transplantation and Hepatobiliary Surgery., Springer, Cham 2020.

[4] L. Sensoy, *A review on the food digestion in the digestive tract and the used in vitro models*, Current Research in Food Science, Volume 4, pp 308-319, 2021.

[5] Y. Nakai, T. Sato, R. Hakuta, K. Ishigaki, K. Saito; T. Saito; N. Takahara, T. Hamada, S. Mizuno, Kogure, H. Tada, H. Isayama, K. Koike. *Management of Difficult Bile Duct Stones by Large Balloon, Cholangioscopy, Enteroscopy and Endosonography. Gut Liver.,Volume. 14, pp.297-305, 15 May 2020.

[6] Lübbe, J. A. *Aspects on advanced procedures during endoscopic retrograde cholangiopancreatography (ERCP) for complex hepatobiliary disorders*, Available from ProQuest Dissertations & Theses Global, 2021. Retrieved from https://www.proquest.com/dissertations-theses/aspects-on-advanced-procedures-during-endoscopic/docview/2493852328/se-2?accountid=53351.

[7] Venkatesh, S.K., Welle, C.L., Miller, F.H. et al. *Reporting standards for primary sclerosing cholangitis using MRI and MR cholangiopancreatography: guidelines from MR Working Group of the International Primary Sclerosing Cholangitis Study Group*. Eur Radiol (2021). https://doi.org/10.1007/s00330-021-08147-7.

[8] H.B. Rao, A.K. Koshy, S. Sudhindran, et al. *Paradigm shift in the management of bile duct strictures complicating living donor liver transplantation*. Indian J Gastroenterol .Volume. 38,p.p 488–497, 2019.

[9] A.P.M Matton, Y. Vries, L.C Burlage, et al. *Biliary Bicarbonate, pH, and Glucose Are Suitable Biomarkers of Biliary Viability During Ex Situ Normothermic Machine Perfusion of Human Donor Livers*. Transplantation. Volume. 7, p.p 1405-1413, 2019.

[10] D. Kursat, C.P Ahmet, Hasan Bostanci; S. Mustafa. *Clinical characteristics and treatment approaches in patients with post-cholecystectomy syndrome due to remnant gallbladder*. Annals of Medical Research. March 2019,Volume. 7,p.p 1172-1177,.

[11] P. Rawla, T. Sunkara; K.C Thandra, A. Barsouk. *Epidemiology of gallbladder cancer*. Clin Exp Hepatol., Volume. 2, p.p 93-102, 2019.

[12] S. Saurabh; B. Green. *Is hyperkinetic gallbladder an indication for cholecystectomy?*. Surg Endosc, Volume 33, p.p 1613–1617, 2019.

[13] B. Nader, I. Yervant; R.M Thomas; V. Kia; K.A. Mouen. *Endoscopic ultrasound-guided cholecystoduodenostomy for acute cholecystitis with removal of large (missed) cystic duct stones*. Endoscopy. Volume 12, 2019.

[14] R. Emmanuelle, E. Vitamin. *Bioavailability: Mechanisms of Intestinal Absorption in the Spotlight*. MDPI: Antioxidants. Volume. 6, pp. 95, 2017.

[15] Hepato Pancreatico Biliary Cancer. 14 August 14, 2021, Retrieved from: https://www.prolifecancercentre.co.in/hepato-pancreatico-billiary-cancer/

[16] N. Mehta, A.T. Strong, T. Stevens, et al. *Common bile duct dilation after bariatric surgery*. Surg Endosc, Volume. 33, p.p 2531–2538, 2019.

[17] Saluja, S.S., Varshney, V.K., Bhat, V.S. et al. *Management of Obscurely Dilated Common Bile Duct with Normal Liver Function Tests: A Pragmatic Approach*. World J Surg 45, 2712–2718 (2021). https://doi.org/10.1007/s00268-021-06175-4.

[18] T. Inomata, K. Nakaya et. al., *Evaluation of the usefulness of cystic duct three-dimensional computed tomography with non-contrast for before laparoscopic holecystectomy and endoscopic transpapillary gallbladder drainage in comparison to magnetic resonance cholangiopancreatography*, Journal of Medical Imaging and Radiation Sciences, Volume 52, Issue 2, Pages 248-256, 2021.

[19] C Saravanan. *Color Image to Grayscale Image Conversion, 2010 Second International Conference on Computer Engineering and Applications*. Bali Island. 2010.

[20] K.K Omeed. *PRODUCE LOW-PASS AND HIGH-PASS IMAGE FILTER IN JAVA 2014*. International Journal of Advances in Engineering & Technology, Volume. 7, p.p 712-722, 2014.

[21] K. Zhang, W.Y Zuo, Chen, D. Meng, L. Zhang. *Beyond a Gaussian Denoiser: Residual Learning of Deep CNN for Image Denoising*. IEEE Transactions on Image Processing, Volume. 26, No 7, pp. 3142-3155, Feb 2017.

[22] TF Chan, LA Vese. Active contours without edges. IEEE Trans Image Process Vol. 10,No 2, pp 266–277, 2017.

[23] Domingues, G.Pereira; P. Martins; H. Duarte; J. Santos; P. H Abreu. *Using deep learning techniques in medical imaging: A systematic review of applications on CT and PET*. Artif. Intell Rev, pp. 1-68, Nov. 2019.

[24] Zhen-Jie Yao, Jie Bi and Yi-Xin Chen. *Applying Deep Learning to Individual and Community Health Monitoring Data: A Survey*. International Journal of Automation and Computing, vol. 15, no. 6, pp. 643-655, 2018 doi: 10.1007/s11633-018-1136-9

[25] P.C. Tripathi, S. Bag. CNN-DMRI: *A Convolutional Neural Network for Denoising of Magnetic Resonance Images*. Pattern Recognition Letters, Volume. 135, PP 57-63, 2020.

[26] B. Theek; T. Nolte; D. Pantke; F. Schrank; F. Gremse; V. Schulz; F. Kiessling. *Emerging Methods in Radiology*. Radiologe, 2020.

[27] T. Mitchell. *Machine learning*. Singapore: McGRAW-HILL, 1997.

[28] F. T. Wang, L. Yang, J. Tang, S. B. Chen, X. Wang. *DLA+: A light aggregation network for object classification and detection*. International Journal of Automation and Computing . http://doi.org/10.1007/s11633-021-1287-y doi: 10.1007/s11633-021-1287-y

[29] H. Migdady. *Boundness of a Neural Network Weights Using The Notion of a Limit of a Sequence, International Journal of Data Mining & Knowledge Management Process (IJDKP)*. Volume. 4, No. 3, May 2014.

[30] H. Migdady; Y. Talafha; H. Alrabaiah. *Controlling the Behavior of a Neural Network Weights Using Variables Correlation and Posterior Probabilities Estimation*. IOSR Journal of Computer Engineering (IOSR-JCE), Vol. 16, Issue 3, Ver. II, PP 36-41, May 2014.

[31] Migdady. *A features extraction wrapper method for neural networks, with application to data mining and machine learning.* Dissertations & Theses – Gradworks.

[32] M. Mazurowski; M. Buda; A. Saha; Bashir M. *Deep learning in radiology: an overview of the concepts and a survey of the state of the art*. 2018.

[33] A. Benou; R. Veksler; A. Friedman; Raviv. Ensemble of expert deep neural networks for spatio-temporal denoising of contrast-enhanced MRI sequences. Medical Image Analysis,Volume. 42, p.p 145–159, 2017.

[34] D. Jiang; W. Dou; L. Vosters; X. Xu; Y. Sun; Tan. *Denoising of 3D magnetic resonance images with multi-channel residual learning of convolutional neural network*. arxiv.org, 2017.

[35] W. Yang; Y. Chen; Y. Liu; L. Zhong; G. Qin; Z. Lu. *Cascade of multi-scale convolutional neural networks for bone suppression of chest radiographs in gradient domain*. Medical Image Analysis, Volume. 35, p.p 421–433, 2017.

[36] R. Peng; L. Zhang; XM. Zhang, et al. *Common bile duct diameter in an asymptomatic population: A magnetic resonance imaging study*. World J Radiol, Volume. 7, p.p 501-508, 2015.

[37] D. Jiang, W. Dou, L. Vosters, X. Xu; Y. Sun, T. Tan. *Denoising of 3D magnetic resonance images with multi-channel residual learning of convolutional neural network*. arxiv.org, 2017.

[38] J. V. Manjon, Coupe P. *MRI denoising using Deep Learning and Non-local averaging*. E-print:1911.04798, 2019.

[39] S. Hossary; A. Zytoon; M. Eid; A. Hamed; M. Sharaan; A. Ebrahim. *MR cholangiopancreatography of the pancreas and biliary system: a review of the current applications*. Curr Probl Diagn Radiol, Volume. 43, p.p 1-13, 2014.

[40] Ivashchenko O.; Rijkhorst E.; Beek L.; Hoetjes N.; Pouw B.;. Nijkamp J; Kuhlmann K.; Ruers T. *A workflow for automated segmentation of the liver surface*. hepatic vasculature and biliary tree anatomy from multiphase MR images, Magnetic Resonance Imaging, 2019.

[41] Selvalakshmi VM, S. Devi. *Segmentation and 3D visualization of liver, lesions and major blood vessels in abdomen CTA images*. Biomedical Research,Volume. 28, p.p 7206-7212, 2017.

[42] Huynh H.; Karademir I.; Oto A.; Suzuki K. *Computerized liver volumetry on MRI by using 3D geodesic active contour segmentation*. Am. J. Roentgenol, Volume. 202, No. 1, pp. 152, 2014.

[43] Selvaraj E.; Culver E; Bungay H.; Bailey A.; Chapman R.; Pavlides M. *Evolving role of magnetic resonance techniques in primary sclerosing cholangitis*. World J Gastroenterol. Volume. 6, p.p 644-658, 2019.

[44] Kozaka K.; Sheedy SP.; Eaton JE.; Venkatesh SK; Heiken JP. *Magnetic resonance imaging features of small-duct primary sclerosing cholangitis*. Abdom Radiol (NY), 2020.

[45] Joo I, Lee JM, Yoon JH. *Imaging Diagnosis of Intrahepatic and Perihilar Cholangiocarcinoma: Recent Advances and Challenges*. Radiology, Volume. 288, p.p 7-13, 2018.

[46] Khan F.; Stevens-Chase A.; Chaudhry R.; Hashmi D.; Edelman; Weavr D. *Extrahepatic Biliary Obstruction Secondary to Neuroendocrine Tumor of the Common Hepatic Duct*. International Journal of Surgery Case Reports. Volume. 30, P.P 46-49, 2017.

[47] Skoczylas K.; Pawe1as A. *Ultrasound imaging of the liver and bile ducts - expectations of a clinician*. J Ultrason, Volume. 15, p.p 292–306, 2015.

[48] Stephen M.; Philip E.; et. al. *Adaptive histogram equalization and its variations, Science Direct: Computer Vision*. Graphics, and Image Processing, Volume. 39, p.p 355-368, September 1987.

[49] He, Kaiming. *Single Image Haze Removal Using Dark Channel Prior*. Thesis, The Chinese University of Hong Kong, 2011.

[50] Dubok, et al. *Single Image Dehazing with Image Entropy and Information Fidelity*. ICIP. pp. 4037–4041, 2014.

[51] Hundt M, Wu CY, Young M. Anatomy, Abdomen and Pelvis, Biliary Ducts. [Updated 2020 Apr 29]. In: StatPearls [Internet]. Treasure Island (FL): StatPearls Publishing; 2020 Jan-. Available from: https://www.ncbi.nlm.nih.gov/books/NBK459246/.

[52] N. a. A. D. Kapre, "Programming FPGA applications in VHDL." Reconfigurable Computing," Morgan Kaufmann, pp. 129-153, 2008.

[53] S. A. e. a. Alazawi, "Texture features extraction based on GLCM for face retrieval system," Periodicals of Engineering and Natural Sciences, vol. 7.3, pp. 1459-1467, 2019.